\documentclass{article}

\usepackage{arxiv}

\usepackage[utf8]{inputenc} 
\usepackage[T1]{fontenc}    
\usepackage{hyperref}       
\usepackage{url}            
\usepackage{booktabs}       
\usepackage{amsfonts}       
\usepackage{nicefrac}       
\usepackage{microtype}      
\usepackage{lipsum}
\usepackage{graphicx}
\usepackage{cite}
\graphicspath{ {./images/} }

\title{VIEW: A Framework for Organization Level Interactive Record Linkage to Support Reproducible Data Science}

\author{
 Mohammad Karim \\
  Population Informatics Lab\\
  Department of Health Policy and Management\\
  Texas A\&M University\\
  1266 TAMU, College Station, TX 77843 \\
  \texttt{karimm@tamu.edu} \\
   \And
 Mahin Ramezani \\
  Population Informatics Lab\\
  Department of Computer Science \& Engineering \\
  Texas A\&M University \\
  1266 TAMU, College Station, TX 77843 \\
  \texttt{mahin@tamu.edu} \\
  \And

Tenaya Sunbury \\
  Washington State Department of Social and \\Health Services\\
  \texttt{tenaya.sunbury@dshs.wa.gov} \\
  \And
  
  Robert Ohsfeldt \\
  Department of Health Policy and Management\\ Texas A\&M University\\
  College Station, TX 77843 \\
  \texttt{rohsfeldt@tamu.edu} \\
  \And
  
  Hye-Chung Kum \\
  Population Informatics Lab\\
  Department of Health Policy and Management\\
  Department of Computer Science \& Engineering \\
  Texas A\&M University \\
  1266 TAMU, College Station, TX 77843 \\
  \texttt{kum@tamu.edu} \\
}

\begin{document}
\maketitle
\begin{abstract}
\textbf{Objective:}
To design and evaluate a general framework for interactive record linkage using a convenient algorithm combined with tractable Human Intelligent Tasks (HITs; i.e. micro tasks requiring human judgment) that can support reproducible data science. 

\textbf{Materials and Methods:}
Accurate linkage of real data requires both automatic processing of well-defined tasks and human processing of tasks that require human judgment (i.e., HITs) on messy data. We present a reproducible, interactive, and iterative framework for record linkage called VIEW (Visual Interactive Entity-resolution Workbench). We implemented and evaluated VIEW by integrating two commonly used hospital databases, the American Hospital Association (AHA) Annual Survey of Hospitals and the Medicare Cost Reports for Hospitals from CMS.

\textbf{Results:}
Using VIEW to iteratively standardize and clean the data, we linked all Texas hospitals common in both databases with 100\% precision by confirming 78 approximate linkages using HITs and manually linking 28 hospitals using HITs.

\textbf{Discussion:}
Similarities in hospital names and addresses and the dynamic nature of hospital attributes over time make it impossible to build a fully automated linkage system for hospitals that can be maintained over time. VIEW is a software that supports a reproducible semi-automated process that can generate and track HITs to be reviewed and linked manually for messy data elements such as hospitals that have been merged.

\textbf{Conclusion:}
Effective software that can support the interactive and iterative process of record linkage, and well-designed HITs can streamline the linkage processes to support high quality replicable research using messy real data.
\end{abstract}


\section{Introduction}
Secondary use of large existing databases for research is increasingly common. The important characteristics of such data are that (1) an extensive amount of data exists on the population served, (2) data are continuously generated, (3) data change over time as programs evolve and originate from multiple sources, and (4) data have varied levels of validity with data directly required for operations being the most valid. These represent the “four Vs” of big data: volume, velocity, variety, and veracity, respectively\cite{kum2013social}. Using big data to extract valuable information requires a tractable and reproducible data processing pipeline. A critical step in the data processing pipeline is data integration. Often called record linkage or entity resolution, it presents a challenge when there is no common, error-free, unique identifier with which to identify records across databases pertaining to the same real-world entities.  

Much has been published about automatic record linkage of person level data\cite{bradley2010health, dusetzina2014overview, elmagarmid2006duplicate, fellegi1969theory, joffe2014benchmark, kum2014privacy, winkler2006overview, etheredge2007rapid, friedman2010achieving, kum2009supporting, antonie2014tracking, zhu2015conduct}. This paper contributes to the literature by documenting the iterative process of developing a linkage algorithm and its application to hospital level data. Documenting the process of linkage is important because researchers often rely on manual, ad hoc tools for data integration due to the lack of standardized approaches or appropriate software. Non-transparent record linkage is a major issue in replicable research.

We present a systematic framework, VIEW (Visual Interactive Entity-resolution Workbench), for incrementally developing a tractable algorithm to link organization-level data. VIEW can be used to develop a well-documented semi-automated process for linking two or more hospital datasets with no common identifiers. We evaluate VIEW in a project that required the development and maintenance of a comprehensive hospital database across five different data sources with timely updates each year.

\section{Significance}
Constructing useful measures for secondary data analysis to answer broad questions often requires the integration of data from multiple systems. For example, our project had to integrate data from five sources, which used a total of four independent identifiers for the providers, the Texas Provider ID (TPI), the National Provider ID (NPI), the Medicare provider ID, and a facility ID (FID). In this paper, we only discuss the process for building a crosswalk from the MedicareID to the FID, which demonstrates the process best. The other linkages were conducted using VIEW in similar ways.  

A major challenge in integrating hospital level data is that hospitals are not static entities but evolve over time (i.e., mergers, closings, name changes, address changes). Thus, maintaining a clean identifier system for all providers over time is challenging. To further complicate this issue, there may be multiple identifiers (i.e., federal, state, and local) used for providers often requiring a system to build a crosswalk between different identifiers when combining data from heterogeneous systems. As a result, there is a pressing need to develop a reproducible process for standardizing and integrating multiple sources of provider level data. There are a variety of applications for such integrated data, including pay-for-performance programs and public reporting, as well as organizational-level quality assurance and performance tracking using big data\cite{newcombe1959automatic, gomatam2002empirical, boscoe2011building}.

\section{Background}
The most common methods for linking individual-level data are probabilistic and deterministic record linkage\cite{bradley2010health, dusetzina2014overview, elmagarmid2006duplicate, fellegi1969theory, joffe2014benchmark, kum2014privacy, winkler2006overview, etheredge2007rapid, friedman2010achieving, kum2009supporting, antonie2014tracking, zhu2015conduct}. The probabilistic method scores a statistical probability of two records being a ‘true’ link based on a model developed typically using training data. Even though there are many different probabilistic methods in statistics and machine learning that currently investigate how to best develop the model given the data, the researcher must still determine two thresholds to group linkages into match, uncertain, or non-match once the data have been scored\cite{dusetzina2014overview}. In comparison, deterministic methods are rule-based, where the researcher specifies the rules under which the two records are considered a match (e.g., pairs that have exact match on name and address), uncertain (e.g., pairs with approximate match on name or address), or non-match (e.g., all other pairs). Often a stepwise approach is used to build the rules\cite{kum2009supporting, antonie2014tracking}.  
  
Probabilistic methods tend to work better on complex data at the cost of less interpretable models. In comparison, simple deterministic methods are easier to implement and communicate when the linkage task is relatively simple, as in the case of linking hospitals. The quality of matching results are comparable for both deterministic and probabilistic methods as long as the process for linkage is well developed\cite{antonie2014tracking, zhu2015conduct}. More importantly, data standardization and cleaning is important in both approaches but also very difficult to do top-down based on theory\cite{randall2013effect}. VIEW includes methods to quickly standardize only the regularities in a given dataset with a bottom-up approach using the data at hand.

To overcome the limitations of automatic algorithms in addressing real world problems\cite{kopcke2010evaluation}, there has been increasing interest in interactive record linkage that better document the human interaction during the linkage process\cite{kang2008interactive, shen2016nameclarifier}. In particular, we present how to use well defined Human Intelligent Tasks (HITs; i.e. micro tasks requiring human judgment), to design effective human machine systems for record linkage. Using HITs is common for processing big data because most tasks require both automatic processing of well defined tasks and human processing for tasks that require judgment\cite{larson2012community}. The importance of human interaction in linkage is demonstrated well in Bronstein et al.\cite{bronstein2009issues} where pregnancies from Medicaid data were linked to birth records via 11 manual steps. There were multiple uncertainties that needed human decisions to attain an overall match rate of 87.9\%. With no human interaction, the match rate would be much lower. Ultimately, the goals of any approximate linkage method should include: 1) setting the match threshold conservatively to avoid the false matched pairs, 2) setting the potential match threshold liberally so all missed true matches are in the uncertain matched pairs and can be recovered during the manual resolution phase, and 3) keeping the number of uncertain pairs (i.e. HITs) to be reviewed manually at reasonable levels.

\section{Data}
The main database is the 2013 provider ID information file that comes with the Hospital Form 2552-10 on the CMS website\cite{cms2014} containing the MedicareID, the name, and address of all providers (N=606 for Texas). To this database, we linked the Texas Annual Survey of Hospitals from 2008 to 2013, which uses the FID. It is a mandatory hospital survey administered by Texas Department of State Health Services working in collaboration with the American Hospital Association and the Texas Hospital Association\cite{dshs}. Some hospitals had multiple values for provider names in the survey because names change over time and both the legal business name and DBA (doing business as) name were available. Hence, there were a total of 800 different names that represented the 664 unique providers.

\section{Method}
\begin{figure}
  \centering
  \includegraphics[width=\linewidth]{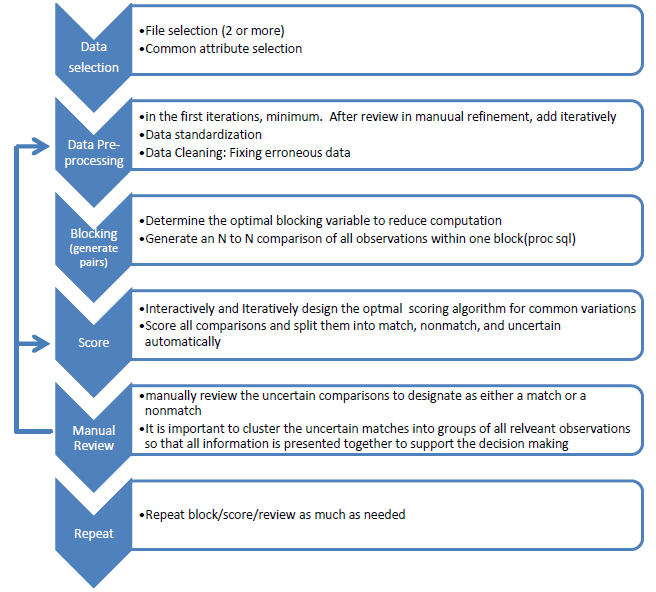}
  \caption{VIEW: Framework for iteratively developing a record linkage algorithm.}
  \label{fig:fig1}
\end{figure}

We first describe methods for measuring linkage quality used throughout the paper. Then, we follow with a presentation of the six core steps of the proposed human machine process (Figure \ref{fig:fig1}) and demonstrate each step using our example linkage study.

\subsection{Measuring Linkage Quality}
The main quality measures in linkage are recall, aka sensitivity, and precision (equation (\ref{1}) and (\ref{2})). Often, the application will determine the balance between recall and precision. In general, setting stringent criteria will result in high precision and low recall whereas looser criteria will start to introduce incorrect matches reducing precision while increasing recall. However, this is not a direct relationship, and carefully building more complex models can increase recall without much reduction in precision. We only report recall in this paper because precision was 100\% in our application.

\begin{equation}\label{1}
    Recall = Sensitivity = \frac{the\, number\, of\, correct\, linkages\, found}{the\, number\, of\, all\, true\, linkages\, that\, exist}
\end{equation}

\begin{equation}\label{2}
    Precision = \frac{the\, number\, of\, correct\, linkages\, found}{the\, number\, of\, total\, linkages\, found}
\end{equation}

\subsubsection{Step 1: Data Selection}
The first step is to select the files to build the crosswalk and then to select the common attributes to be used in the matching process. Good attributes to use are variables that tend to be recorded consistently and have high distinguishing power (i.e. many unique values). For example, with only two possible values, type of hospital (i.e. public or private) is a low power variable. In comparison, with mostly unique values, name is a high power variable. However, names tend to have a lot of variation for the same entity, which decrease its usefulness. The discriminatory power of identifiers can be quantified using the Shannon entropy\cite{shannon1948mathematical}. In our linkage, the common data attributes were provider name, city, zip code, and street address. We dropped city because it had similar information as zip code, and the more granular data numerically coded was better.  

\subsection{Step 2: Data Standardization and Cleaning}
Variation in the way that attributes are represented across data used to link hospitals can result from different coding methods (e.g. use of uppercase versus lowercase), the dynamic nature of the underlying attributes (e.g. renaming a hospital after a change in ownership), erroneous data (e.g. typos), or missing data.  Standardization of common data elements both in terms of formats (capitalization) and values (i.e. street to st) reduces the unnecessary variations in the data and significantly improves automatic linkage. Numerically coded attributes using the same coding scheme work best. For example, zip code works well for linking organizations because it has high distinguishing power as well as low variation in common values. Nonetheless, developing well coded variables is time intensive and often linkage is carried out on raw data without the common coding by carrying out approximate matches.

The most efficient method to standardizing and cleaning the data is to set up a data processing pipeline to easily add in standardization rules iteratively over time as problems are discovered in the data. Setting up such a framework for processing big data is critical as it is difficult to know up front all the issues with any given dataset. Thus, as researchers encounter different issues in the data, the ability to go back and add rules to clean the data, then easily repeat the steps is essential to working efficiently with big data in a tractable manner. In record linkage, this means that in the first iteration, there are likely to be no data cleaning or standardization rules because the researcher does not know the issues in the data yet. Such rules will be developed and incorporated in subsequent iterations.  

Using computer code to automate data cleaning and standardization has several advantages compared to manually editing the data. First, if the process is automated, then work will not be lost if rules need to be revised or deleted in subsequent iterations. In addition, the computer code serves as documentation of what was done. Such documentation is important for making research reproducible. And finally, using an automated process makes it simple to retract any steps that are later detected as incorrect during the process of working with the data.

Two ways of effectively standardizing provider names quickly are to drop frequently used words (e.g. hospital) and to replace terms that are frequently abbreviated (e.g. center, ctr, cntr) with a standard set of consistent abbreviations.  VIEW provides a module that produces the frequent word list. Detecting the commonly used abbreviations occurs iteratively during manual review of uncertain and non-matched records.  

\begin{figure}
  \centering
  \includegraphics[width=\linewidth]{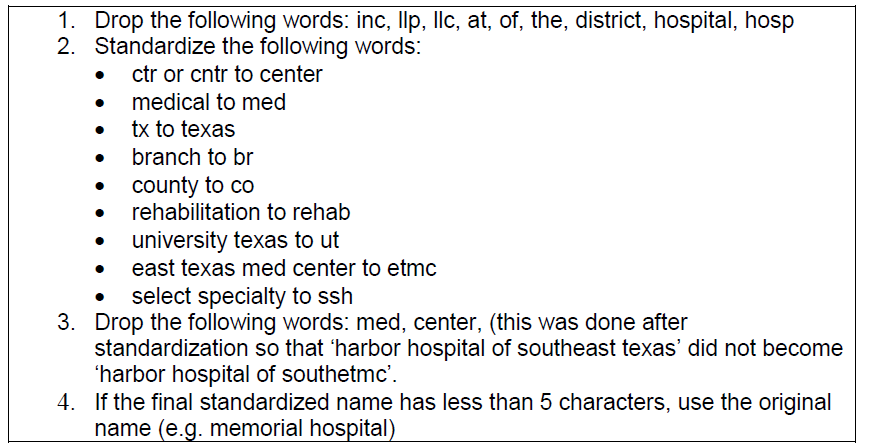}
  \caption{Final Name Standardization Algorithm.}
  \label{fig:fig3}
\end{figure}

In our linkage, basic standardization (i.e. using only lower case and removing all special characters) improved the recall rate to 51\%.  Figure \ref{fig:fig3} and Table \ref{t1} are the final standardization we used for name and address after multiple iterations. Note that the order of the standardization matters. It was also important to have the last step where we use the original name when the standardized name becomes null (e.g. memorial hospital). Using these standardizations, exact match on standardized names improved the recall rate to 67\%. 

\begin{table} \label{t1}
 \caption{Final Address Standardization}
  \centering
  \begin{tabular}{|c|c|c|}
    \hline &&\\ 
    \textbf{Original word} & \textbf{Standardized to}	& \textbf{Ignore for approximate match} \\ \hline 
    lane & ln &	X \\ \hline
    street & st & X \\ \hline
    boulevard, boulevard & blvd & X \\ \hline
    road & rd & X \\ \hline
    circle & cir & X \\ \hline
    drive &	dr & X \\ \hline
    avenue & ave &	X \\ \hline
    loop & lp &	X \\ \hline
    ctr, cntr & center & \\ \hline
    $3^{rd}$, $2^{nd}$ &	third, second respectively & \\ \hline 
    highway, freeway, parkway & hwy, fwy, pkwy respectively & \\ \hline 
    north, south, east, west & n, s, e, w respectively & \\ \hline	
  \end{tabular}
\end{table}

\subsection{Step 3: Blocking}
The full comparison space is the Cartesian product of the two datasets being linked, majority of which are non-matches (e.g, 606*800=484,800 comparisons in our example). To reduce the search space, one or more blocking variables are used to compare only records that share the attribute. Blocking can introduce problems when there are data errors or missing values in the blocking variable because the correct comparisons cannot be made. Thus, it is common to use a multi-pass blocking algorithm to recapture those comparisons that are permanently lost in the first pass. Clearly, the blocking variable has direct impact on performance in terms of time and quality.  

In our study, zip code is the best blocking variable because it will break up the data into small number of hospitals in each zip code to be matched up with no missing data (Table \ref{t2}). Blocking on zip code reduced the number of comparisons from 484,800 to only 1,752. However, there were 37 entities with incorrect zip code, which had to be dealt with in the second pass.

\begin{table}\label{t2}
 \caption{Search Space}
  \centering
  \begin{tabular}{|c|c|c|}
    \hline &&\\ 
     & \textbf{Medicare data}	& \textbf{AHA Survey data} \\ \hline 
    Unique \# of zip codes & 407 & 418 \\ \hline
    Mean \# of hospital per zip code & 2.2 & 3.0 \\ \hline
    Max \# of hospital per zip code & 10 & 13 \\ \hline
  \end{tabular}
\end{table}

\subsection{Step 4: Scoring}
The next step is the pairwise scoring of all pairs within each block. In the first iteration, only simple standardization and a simple scoring system (i.e., if all common attributes match exactly classify as a match, if at least one attribute match approximately classify as uncertain, otherwise classify as nonmatch) is used. For most problems, this simple setup will result in a low match rate and a large number of uncertain matches. In the first few iterations, you are scanning both the uncertain and nonmatch groups for regular patterns in the two dataset that you need to either standardize or use for scoring a pair. As you spot them, you add the standardization code and the scoring code to be more complex and rerun until there are no more improvements you can do automatically. The goal is to iteratively develop both the standardization and scoring algorithm to capture regularities in the data automatically as true matches and reduce the uncertain group to include only the difficult cases that require human judgment. In addition, you should be reviewing the nonmatch group to confirm that these are indeed nonmatches. Typically, you will spot required standardizations (e.g. using same abbreviations such as East Text Medical Center to ETMC) in the nonmatch group in the beginning.  

Probabilistic record linkage methods develop statistical models for automatic scoring using training data that have been manually labeled. Then the researcher determines the two thresholds for match, uncertain, and nonmatch in the final score. However, for reasonable sized data, using simple rule based deterministic scoring methods is more tractable and interpretable and works comparably. Deterministic methods are also easier to control precisely what you group for manual review versus automatic linkage.

\begin{figure}
  \centering
  \includegraphics[width=\linewidth]{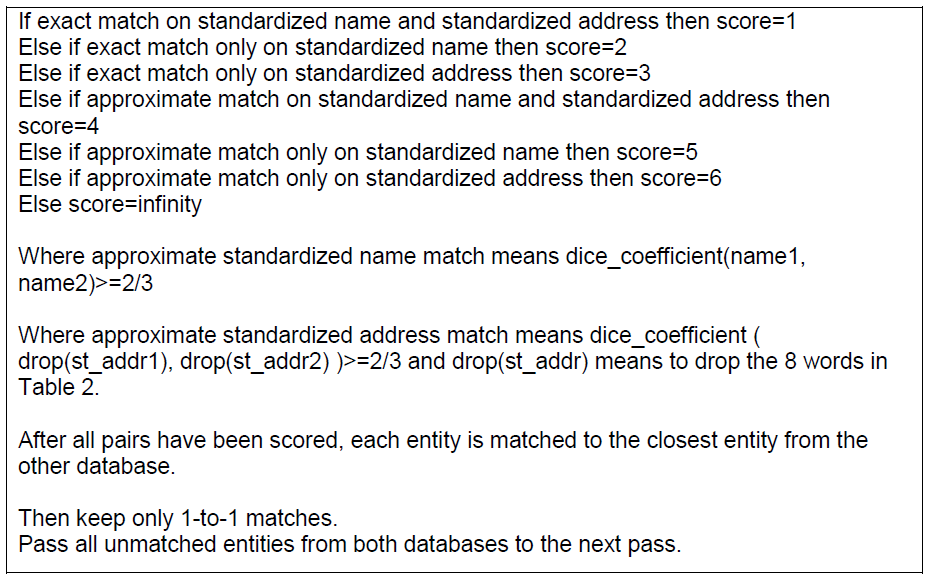}
  \caption{Scoring Algorithm.}
  \label{fig:fig4}
\end{figure}

In our linkage, we allowed for deterministic approximate matching on both name and address as detailed in Figure \ref{fig:fig4} which further improved the recall to 88\%.  The algorithm, builds a non-directional graph with each entity in one database connected to all other entities in the other database in the same zip code. Then each link is scored on a priority of 1 to 6 based on the similarity of names and addresses. If both name and address does not match at all, the link is deleted. As a final step, for each pair of entities only the closest link is kept. That is if there are two names, and one matches exactly, only the exactly matched name is considered. Then the main criteria used for automatic linkage was to only allow for linkage that were 1-to-1 linkages in the remaining graph. Any linkages that resulted in more than 1 mapping was kicked out for next iteration. The importance of 1-to-1 mapping criteria for automatic linkage is discussed later.

VIEW makes it easy to add in customized SAS code for approximate match on each variable and provide macros for counting the number of common words and the dice coefficient of common words. The dice coefficient is commonly used to measure the similarity of sets and is defined as two times the number of common words over the total number of words. The threshold of 2/3 for the dice coefficient can account for addresses having fewer details. For example, the dice coefficient for ‘865 deshong’ and ‘865 deshong 5th floor’ is 2*2/6=2/3. Of the 521 matches made, 40\% were exact match on std\_name and std\_addr, 30\% were exact match on only std\_name, 22\% were exact match on std\_addr, with the remaining 8\% being an approximate match of some kind. These 41 approximate matches were generated as HITs and confirmed manually. There were two in the HITs that required further investigation outside the databases to confirm as a correct match.

\subsection{Step 5: Second Iteration}
You can add as many block/score/review pass as needed to recapture any matches not compared in a particular blocking pass. Typically, in subsequent passes you are only processing data that have not been linked in the previous pass. This kind of divide and conquer method is very effective for working with big data.

In our linkage, after the first pass blocking on zip code and allowing for approximate match on standardized name and address for 1-to-1 matches, there were 85 MedicareIDs and 171 FIDs that were not matched. Given the small numbers, we ran a second pass without blocking by linking any records that were an exact match on standardized name (43 matches) or standardized address (12 matches). All except 3 matches were 1-to-1 matches. 13 were both name and address match while 39 only matched on one. We generated the 39 as HITs to be confirmed for accuracy. All except two of these were matches with different zip code due to an error in one of the datasets. This improved the recall to 97\%.

\subsection{Step 6: Manual Review}
Once automatic linkage is developed, the program should generate three separate outputs, one for confirmed automatic matches, one for the potential matches, and one for any entities that did not link to anything. The potential matches are HITs that are output into an excel file. These HITs are manually resolved into another excel sheet so that human judgment can be incorporated back into the process as well as documented. Manual refinement can occur at the end of any block/score pass.

We had one manual review step at the end. After the second pass, we had 33 MedicareIDs and 105 FIDs that were still not matched. These were output as HITs to be matched manually. We easily found 28 matches manually. Of the 6 remaining, 3 were duplicate records and 3 were those that did not participate in the survey.

\section{Results}
\subsection{Software for Linking Data}
We have developed and released VIEW under GNU license to facilitate replicable methods in linking data. VIEW is a set of general SAS macro codes that implements the record linkage described above for deterministic methods. It can be easily extended to perform probabilistic methods as well. Researchers can specify and control many aspects of the linkage such as how to standardize and score data by adding customized SAS code to designated files. As in our survey data, often entities have multiple names. Thus, VIEW provides an easy mechanism for properly managing multiple rows per entity so that if any of the names match the correct linkage is made by keeping the primary ID the same. More details can be found on the VIEW website\cite{view}.

\subsection{MedicareID to FID}
\begin{figure}
  \centering
  \includegraphics[width=\linewidth]{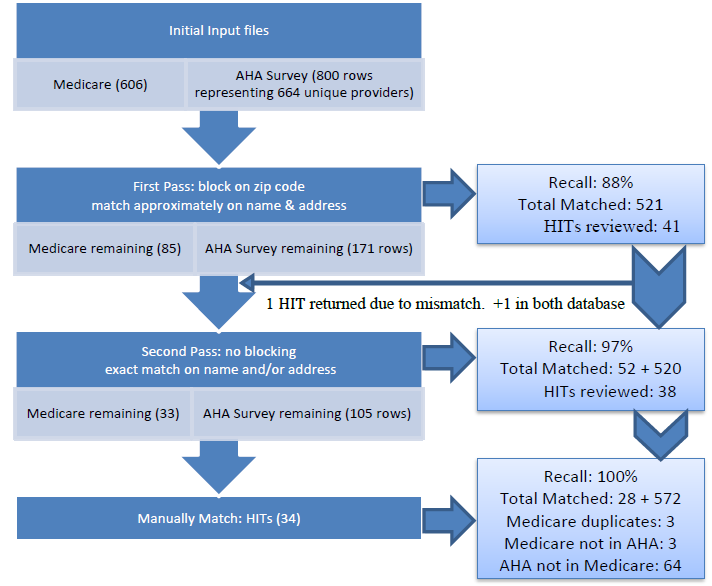}
  \caption{Medicare ID to FID linkage process.}
  \label{fig:fig5}
\end{figure}

Figure \ref{fig:fig5} depicts the full process for using VIEW to link the MedicareID to FID After using VIEW to iteratively standardize and clean the data, we (1) automatically linked 493 providers, both exact and approximate match, (2) manually linked 28 providers, (3) confirmed valid no links for 6 providers, and (4) confirmed approximate links for 78 providers and no link for 1 approximate link. Most of the linkages that were manually linked using HITs could never be coded as automatic matching due to the complexity and insufficient information in the database. We had to use additional information found on the internet to confirm the links. To obtain 100\% precision, this application used conservative criteria for automatic matching leading to more manually reviewed HITs.

\subsection{Discussion}
\subsection{1-to-1 link as criteria for automatic matching}
Of all possible matches, determining the critical conditions that confirm an automatic match is important but difficult. In our study, we found that a clean 1-to-1 link can confirm a match automatically whereas links that have multiple matches in the database was a signal for potential issues with similar standardized names among different entities or multiple providers located on the same street. This is because the chances of a provider in each of the databases being an exact 1-to-1 match purely by chance is negligible given the possible range of values\cite{antonie2014tracking}. Thus using the 1-to-1 match criteria is a good rule for protection against false matches resulting from reducing too much variation through standardization such that two different providers have the same standardized name. These errors show up as an N-to-1 match, and need human judgment. This is intuitive in that, in sparse data space linkages are easy where as in dense data spaces (i.e., many entities with similar names) linkages require more attention.  

\subsection{Subtle differences in entities}
There are differences in how an entity is defined in different hospital ID systems. For example, the FID used in the survey data is closely associated with the hospital licensing number. Any change made to the provider license over time is reflected in the FID and the FID is managed manually by state staff to ensure high quality data. Thus, we can track changes in names, addresses, or closures over time by tracking the same FID. This means that same health systems can have one or more FIDs depending on how they are licensed. On the other hand, the MedicareID is for billing, and an entity is a Medicare provider, which may or may not correspond to their licensed structure. The most common ID system being used for hospitals, NPI, often have multiple IDs for one health system making entity resolution very difficult. In our linkage, the Medicare data had two entities from one health system with different zip codes that were adjacent (walking distance), but on the same street (different street number). In comparison, the survey data only had one entity from the same hospital system. This was one of the linkages we had to investigate beyond the data at hand. Based on the number of licensed beds in the survey and their website, we concluded the correct linkage.

\subsection{Same name or address for different entities}
A small number of providers have the same name even when they are different entities. Most of these are hospitals in the same system in different locations with separate licenses. In the survey data, we had 6 providers, which had the same name for multiple entities. To differentiate them, we added the city name to the provider name. In addition, the standardization we used made both ‘University Medical Center’ and ‘University Hospital’ become ‘University’. Thus, exact matching on names, can lead to 1 erroneous match. However, since these were hospitals in two different zip codes, these providers were properly matched to ‘University Medical Center: Lubbock’ and ‘University Health System’ respectively when we blocked on zip code before scoring.  

There were many providers on the same street but in combination with street number and provider name, these did not cause problems. There was a pair of providers that had both a psychiatric license and an acute care license at the exact same address in the survey. There were also two in the Medicare data, which we could match up manually. ‘SSH South Dallas’ is located on the 4th floor of Methodist Charlton Medical Center, which also caused confusion and required human judgment.

\section{Limitations}
The manual work in record linkage is inherently dependent on the data to be linked. Both errors in data as well as gaps in how the same entity is represented in the different databases requires iteratively interacting with the data, detecting these patterns, and coding these patterns into the process to clean the data. VIEW is setup to make this process more efficient and tractable, but cannot replace the required hard work for replicable research.

\section{Conclusion}
Similarities in provider names and addresses, the dynamic nature of hospitals over time, and the subtle differences in entities make it impossible to build fully automated hospital linkage system. However, manually managing data linkages for even a small number, particularly over time, is inefficient, could lead to human error, and difficult to replicate. Thus, effective software that can support the interactive and iterative process of record linkage and well-designed HITs streamline data linkage processes supporting high quality replicable research using big data.

\bibliographystyle{unsrt}  
\bibliography{references}  

\end{document}